\documentstyle[aps,prb,twocolumn,psfig,floats]{revtex}

\begin{document}
\draft

\title{\Large\bf Electronic and structural properties of  vacancies on
  and below the GaP(110) surface}

\author{G. Schwarz, A. Kley, J. Neugebauer, and M. Scheffler}

\address{
Fritz-Haber-Institut der Max-Planck-Gesellschaft,\\ Faradayweg 4-6,
D-14195 Berlin-Dahlem, Germany
}
\date{\today}

\maketitle

\begin{abstract}
We have performed total-energy density-functional calculations using
first-principles pseudopotentials to determine the atomic and electronic
structure of neutral surface and subsurface vacancies at the GaP (110) surface.  
The cation as well as the anion surface vacancy show a pronounced inward
relaxation of the three nearest-neighbor atoms towards the vacancy while the
surface point-group symmetry is maintained.  
For both types of vacancies we find a singly occupied level at midgap.
Subsurface vacancies below the second layer display essentially the same
properties as bulk defects.
Our results for vacancies in the second layer show features not observed for
either  surface or bulk vacancies: Large relaxations occur and both defects 
are unstable against the formation of antisite vacancy complexes.
Simulating scanning tunneling microscope pictures of the different vacancies,
we find excellent agreement with experimental data for the surface vacancies
and predict the signatures of subsurface vacancies.
\end{abstract}

\pacs{PACS numbers: 68.35.Dv, 71.15.Mb, 73.20.Hb}

\section{Introduction}
Native defects at and near surfaces are commonly invoked in explaining various
electrical and structural properties of surfaces and interfaces such as
Fermi-level pinning at clean and metal covered 
surfaces.~\cite{spic:79-2,spic:80-2,wal:87}  
Also properties such as the surface chemical reactivity,
nucleation processes, or diffusion at surfaces and interfaces~\cite{wal:89}
depend to a large extent on the presence and properties of native defects. 
Motivation for the study of surface defects also arises from the investigation
of bulk impurities:
Cleaving crystals in ultrahigh vacuum offers the opportunity to study defects
at and near the cleavage plane.
Previous investigations on dopant atoms,~\cite{john:93e,zheng:94,ebert:96-4}
antisites,~\cite{feen:93} and defect complexes~\cite{ebert:97-1} have shown
that the concentration of such impurities corresponds well to the measured
bulk concentrations.  
However,  it is by no means obvious how the structural and optical properties
of the defects are changed if they are close to or within the surface.
We therefore compare theoretically the properties of defects at and below
surfaces with their corresponding bulk defects in order to interpret the
experimental data measured on cleavage planes. 

The aim of the present paper is to analyze how the properties and behavior
of native defects change with their distance from the surface. 
In particular, we focus on vacancies since these defects have been
experimentally studied in bulk systems~\cite{ken:83,saa:93-1,lain:96} as well
as at the (110) surface.~\cite{cox:90-2,gwo:93,leng:94-1,ebert:94-1,smith:96-1}
The atomic as well as the electronic structure of cation and anion
vacancies in different surface layers and in a bulk environment are
discussed. We also present results concerning the formation energies of
the different vacancies as well as theoretical STM pictures that can be
directly compared to experiment.
We will show that subsurface vacancies are significantly different from
surface or bulk vacancies: 
Vacancies in the second surface layer are unstable against the formation of
antisite vacancy pairs. 

The paper is organized as follows:
First we briefly describe our method.
In Sec.\,\ref{section3} we present the atomic and electronic structure for
gallium and phosphorus surface vacancies. In Sec.\,\ref{section4} these
results are compared with calculations of vacancies below the surface and in
the bulk. 
We discuss further the formation energies and conclude on the abundance in
thermal equilibrium in Sec.\,\ref{section6}. 
Based on the calculated atomic and electronic structure, STM images of the
various surface and subsurface defects are simulated and will be discussed in
Sec.\,\ref{section5}. 
Finally we will summarize our results.

\section{Calculational method}
Our calculations employ density-functional theory using the local-density
approximation (LDA) for the exchange-correlation
functional.~\cite{ca:80,perzu:81} 
Fully separable, norm-conserving pseudopotentials~\cite{fuchs:97} are used to
describe the interaction of electrons and ions. 
The eigenfunctions of the Kohn-Sham operator are expanded in plane
waves~\cite{ihm:79} using a cutoff energy of 8 Ry.  
Tests were performed for the electronic structure using a cutoff up to 15\,Ry.
They did not result in significant changes in the character of the wave
functions or the formation energies of the defects.
For the integration in reciprocal space one special {\bf k}-point in the ($2\times
4$) surface Brillouin zone (point-group $C_{1h}$) is  used.~\cite{monk:76}
With this choice of basis set we find the theoretical lattice constant to be
$5.35$\,\AA{}.  
This value is $1.8$\% smaller than the experimental result,~\cite{data:91} 
which is typical for a III-V semiconductor calculated within the LDA 
(Ref.~\onlinecite{fil:94}) and neglecting zero-point vibrations.

The calculations of the bulk defects are performed in a  64-atom cubic
supercell. 
A 96-atom supercell with six layers of GaP and a vacuum region corresponding
to four layers is used for the simulation of defects on and below the (110)
surface. The surface cell in these calculations has a $(2\times 4)$ geometry
in the $[001]$ and $[1\overline{\rm 1}0]$  directions 
and the defect is created on one side of the slab.
For the calculation of defects, structure optimization is performed to the
first two outermost layers for the surface defects, to the first three layers
for defects in the second surface layer, and to nearest and second nearest
neighbors for all other vacancies.  
In order to quantify the errors due to the finite thickness of our slab and
surface unit cell, we  calculated the gallium surface vacancy also in larger
$(3\times 4)$ and $(2\times 5)$ surface cells as well as for up to eight layers in
the $(2\times 4)$ cell.  
These tests resulted in a variation of defect formation energies less than
50\,meV (Table~\ref{tab01}) while the character and eigenvalues of the defect
states remained almost unchanged. 
Also the simulation of  STM images proved to be well converged within these 
variations of cell size and cutoff energy.

\begin{table}[t]
\caption[]
 {\label{tab01}
Formation energies $E_{\rm f}$ of the gallium surface vacancy ($V_{\rm Ga}$) for
different cell sizes. 
All values are given in eV and under phosphorus-rich conditions ($\mu_{\rm
  P}=0$). The details of the calculation are given in
Sec.~\ref{section6}.\\ 
}
\begin{tabular}{lcccc}
                  &\multicolumn{4}{c}{$V_{\rm Ga}$} \\
surface unit cell &$(2\times 4)$&$(2\times 4)$&$(3\times 4)$&$(2\times 5)$\\
slab layers       & 6           & 8           &  6          & 6 \\
\hline
$E_{\rm f}$       &1.95         &1.97         &2.00         & 1.99  \\
\end{tabular}
\end{table}

\section{Surface vacancies}\label{section3}

\begin{figure}[t]
\centerline{\psfig{file=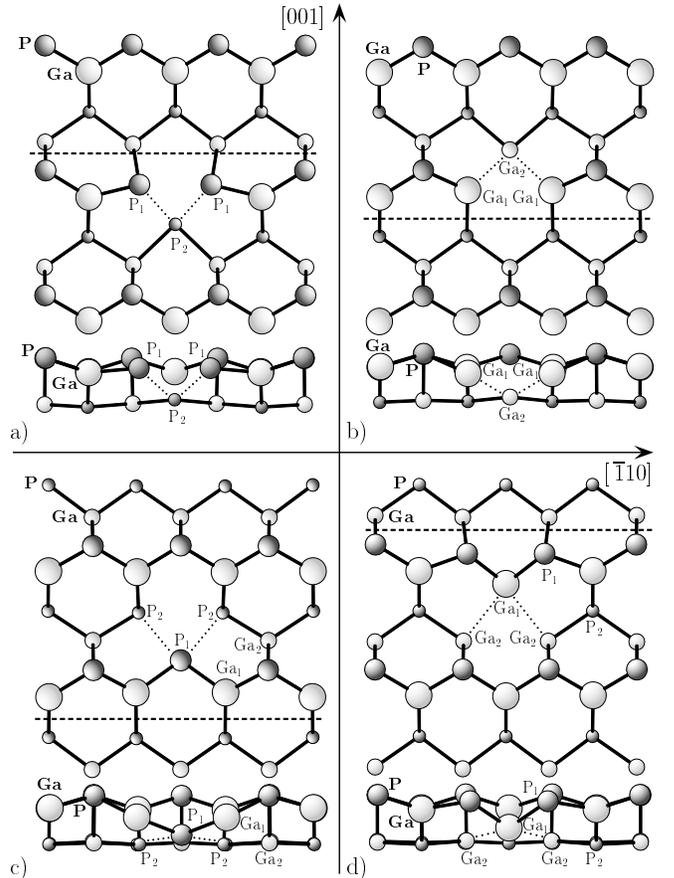,height=11.7cm,clip=t}}
\caption[]
 {\label{fig01}
Atomic relaxations for the defects in the first and second surface
layer. For each  defect the  top and side view of the first two layers are
shown.
Big atoms belong to the surface layer and small ones to the second layer. 
Crystallographic directions are indicated as well as the planes taken for 
the side-view pictures (dashed lines crossing the top view).
(a)~$V_{\rm Ga}$ in the surface layer; 
(b)~$V_{\rm P}$ in the surface layer; 
(c)~$V_{\rm Ga}$ in the second  layer; 
(d)~$V_{\rm P}$ in the second layer.
}
\end{figure}

\subsection{The gallium vacancy}
First let us briefly recall the properties of the defect-free GaP(110) surface,
which has been extensively studied.~\cite{duke:93} 
The point-group $C_{1h}$ of this surface consists of a single mirror
plane in the $[1\overline{\rm 1}0]$ direction and the relaxation of surface
atoms is characterized by a pronounced outwards movement of the anion and an
inward shift of the cation in the surface layer. 
The result is a characteristic buckling of the surface atoms, which is commonly
described by a tilting angle $\omega$. We find this angle to be $27^{\rm o}$
and the height difference between gallium and phosphorus surface atoms to be
$0.56$\,\AA{}, in good agreement with previous theoretical~\cite{alv:91,klep:93}
and LEED data.~\cite{mail:85} 
The relaxation is driven by the lowering of the filled surface state located
at the surface anions.  
As a result of the surface relaxation, the back bonds of the anions are nearly
perpendicular while the cations show a planar bonding configuration indicating
that these atoms are  $p$ and $sp^2$ bonded. 
The gain in surface energy due to relaxation is $650$~meV per surface unit
cell. 

Let us now consider a gallium vacancy $V_{\rm Ga}$ at the surface: Removing
a cation from  the surface layer results in creating dangling orbitals at the
two neighboring phosphorus atoms $\rm P_1$ at the surface and one at the
neighboring atom $\rm P_2$ in the second layer of the slab [Fig.\,\ref{fig01}(a)]. 
The phosphorus atoms $\rm P_1$ in the first layer are twofold and the atom
$\rm P_2$ in the second layer is threefold coordinated. The calculated
equilibrium positions are summarized in Table\,\ref{tab02}.  
It is interesting to note that only the nearest neighbors around the defect
significantly relax. All other atoms remain close to the positions of the
unperturbed surface with a maximum relaxation smaller than  $0.1$\,\AA{}.
The relaxation of the nearest neighbors is characterized by a strong
inward relaxation towards the vacancy. 
The atoms $\rm P_1$ move inwards and lift the surface buckling locally almost
completely. 
We also looked in detail for asymmetric  distortions by starting from various
low-symmetry configurations with equivalent {\bf k}-point sampling. However,
for neutral vacancies the energetically most stable structure maintains the
$C_{1h}$ symmetry of the (110) surface.

\begin{table}
\caption[]
 {\label{tab02}
Relaxations of the nearest-neighbor atoms of the cation ($V_{\rm Ga}$) and
anion ($V_{\rm P}$) surface vacancy. $A_x$ refers to the atom of species
$A$ at the surface ($x=1$) or in the second surface layer ($x=2$).
 $\Delta{\rm [001]}$, $\Delta{\rm [1\overline{\rm 1}0]}$, and $\Delta{\rm [110]}$
 indicate the shifts in the directions illustrated in
 Fig.\,\ref{fig01}; $||\Delta||$ is the total displacement, and
$\alpha$ the bonding angle of the twofold-coordinated surface atoms and of
the bonds in the plane parallel to the surface for the subsurface atoms.\\
}
\begin{tabular}{cc*{4}{r@{.}l}}
     &   &\multicolumn{4}{c}{$V_{\rm Ga}$} &\multicolumn{4}{c}{$V_{\rm P}$} \\ 
     &   &\multicolumn{2}{c}{$\rm P_1$} & \multicolumn{2}{c}{$\rm P_2$} &
     \multicolumn{2}{c}{ $\rm Ga_1$} &\multicolumn{2}{c}{ $\rm Ga_2$} \\ 
\hline
$\Delta{\rm [001]}$       & ({\small \AA}) &$-0$&${61}$&$0$&${49}$&$ 0$&${30}
$&$-0$&${30}$ \\ 
$\Delta{\rm [1\overline{\rm 1}0]}$ & ({\small \AA}) &$
0$&${33}$&\multicolumn{2}{c}{ }  
                &$-0$&${06}$ &\multicolumn{2}{c}{ } \\
$\Delta{\rm [110]}$       & ({\small \AA}) &$-0$&${44}$&$0$&${27}$&$-0$&${29}
$&$ 0$&${14}$ \\ 
 \\
$||\Delta||$                & ({\small \AA}) &$ 0$&${82}$&$0$&${56}$&$
0$&${42} $&$ 0$&${33}$ \\ 
 \\
$\alpha$ & (deg.) & \multicolumn{2}{r}{90.1} & \multicolumn{2}{r}{93.1} &
\multicolumn{2}{r}{120.3} & \multicolumn{2}{r}{98.6}
\end{tabular}
\end{table}

\begin{figure}[t]
\centerline{\psfig{file=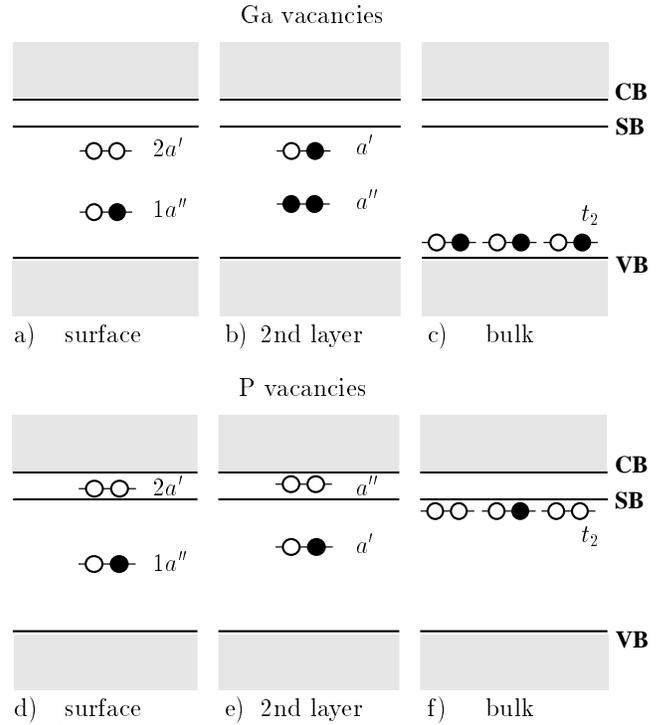,height=9.6cm,clip=t}}
\caption[]
 {\label{fig02}
Single-particle energies for surface vacancies, vacancies in the second surface layer, and vacancies in the bulk. 
Top: results for gallium vacancies; bottom: results for phosphorus vacancies. 
Also indicated are the theoretical values for the valence-band maximum VB
($E=0.0$\,eV), the unoccupied  surface band SB ($1.5$\,eV), and the conduction-band minimum CB ($1.7$\,eV). 
}
\end{figure}

Due to the large relaxation of the phosphorus atoms towards the center of the
vacancy, the distance between these atoms is strongly  reduced. 
We find $2.7$\,\AA{}  between $\rm P_1$ and $\rm P_2$ and $3.1$\,\AA{}  for
the $\rm P_1 P_1$ separation, as compared to $3.8$\,\AA{} for the second
nearest-neighbor distance in bulk GaP or $2.2$\,\AA{} in black
phosphorus.~\cite{gmel:64}   
This indicates that the driving force behind the strong relaxation is the
tendency to form phosphorus-phosphorus bonds.
For the  twofold-coordinated atoms  $\rm P_1$ we find a bond angle of
$90.1^{\rm o}$, which is slightly closer to an ideal  $p$ configuration  than
for the anions at the surface without a defect. 

Figure\,\ref{fig02}(a) summarizes the results for the single-particle energies of
the gallium surface vacancy. 
We find two defect-related states in the fundamental band gap:~\cite{foot:4}
One that we call $1a^{\prime\prime}$ is located $0.5$\,eV above the
valence band, the other called $2a^{\prime}$ is close to the upper surface
band. For the neutral charge state, $1a^{\prime\prime}$ is singly occupied
and $2a^{\prime}$ is empty.  
From analyzing the  wave functions and the local density of states (LDOS), we
find that both $1a^{\prime\prime}$ and $2a^{\prime}$ consist mainly of
$p$-states centered at the three $\rm P$ neighbors.
From this analysis we find further a resonance $1a^{\prime}$ at
about $-1$\,eV also consisting mainly of $p$-states centered at the
three $\rm P$-neighbors. 
The lowest-lying localized level is determined at
about $-8.5$\,eV and can be characterized as back bonds of the three nearest-neighbor atoms. 

The symmetry and nature of these defect states can be qualitatively
understood in terms of a simple model:~\cite{harr:89}
Creating a vacancy results in three dangling orbitals at the two
surface atoms $\rm P_1$ and at the atom $\rm P_2$ in the second
layer. Rehybridization of these orbitals will give rise to three
nondegenerated defect states. 
According to the point symmetry of the (110) surface, these defect states are
either symmetric or antisymmetric with respect to the mirror plane. This is
different from the bulk system, where $T_d$ symmetry gives rise to $a_1$ and
$t_2$ states formed by the dangling orbitals of four nearest-neighbor
atoms.~\cite{bach:81,scheff:81} 
Analyzing the character of defect-related wave functions, we find the
$1a^{\prime}$ state  to be symmetric while the singly occupied state
$1a^{\prime\prime}$  is antisymmetric. The highest level $2a^{\prime}$
is symmetric again.

To our knowledge this is the first  {\it ab initio} calculation of cation
surface vacancies.
Allen, Sankey, and Dow~\cite{all:86} calculated the neutral gallium surface
vacancy in an empirical tight-binding scheme: Neglecting atomic  relaxations
and electronic self-consistency, they found  a singly occupied level $0.5$\,eV
above the valence-band maximum and an empty level at $1.1$\,eV.
These results contrast ours: Without including
structural optimization, we find two nearly degenerate levels above the 
valence-band maximum, but upon atomic relaxation the levels split and take the 
energies shown in Fig.\,\ref{fig02}(a). 

The electronic structure of surface defects can be investigated experimentally
by scanning tunneling microscopy (STM) and scanning tunneling spectroscopy
(STS). 
In STM, the charge states can be estimated by examining defects on $p$- and
$n$-type doped crystals and comparing the 
local surface band bending caused by the charged defects with that of other
charged impurities such as ionized dopants.~\cite{smith:96-1}
In STS, a tunneling spectrum at the defect site gives directly the position of
ionization energies in the band gap as has been done for antisite defects at
the GaAs (110) surface.~\cite{feen:93}
From our results, a first estimate of ionization energies of the defect can be
obtained by the positions of the defect levels in the neutral charge state.
As the gallium surface vacancy has a singly occupied state at midgap, 
we expect at least three charge states \mbox{($+,0,-$)}. While we are not
aware of STS measurements on surface vacancies, this result appears to be in
contrast to STM measurements by Ebert,~\cite{ebert:93-4,ebert:98-1} where 
the gallium surface vacancy was found to be neutral for {\em both} $n$- 
and $p$-type material.
However, it is important to take into account that the charge state of a
surface defect is not determined by the position of the bulk Fermi level but
by the position of the surface Fermi energy. 
Indeed, photoelectron spectroscopy experiments~\cite{reich:97-1} on GaP(110)
indicate that the Fermi level at the surface is pinned near midgap, which
might stabilize the neutral charge state.
Further theoretical and experimental studies are necessary to clarify this issue.

\subsection{The phosphorus vacancy}\label{section3b}
The phosphorus surface vacancy, $V_{\rm P}$, is surrounded by three 
low-coordinated gallium atoms; two in the first layer ($\rm Ga_1$) and one 
in the second layer ($\rm Ga_2$). 
The calculated equilibrium geometry of the defect is summarized in
Table\,\ref{tab02} and Fig.\,\ref{fig01}(b). 
As for the gallium vacancy, the equilibrium structure maintains the $C_{1h}$ 
symmetry of the defect-free surface and atomic relaxations are
mainly restricted to nearest-neighbor atoms. Second nearest neighbors relax
less than $0.1$\,\AA{}.  
However, the inward relaxation of the surface atoms  $\rm Ga_1$ is clearly
less pronounced than for the gallium vacancy.  
The origin of this is the different configuration of surface cations and
anions on the defect-free surface: 
Since surface gallium atoms at the relaxed but unperturbed surface are
located deeper in the surface than phosphorus atoms, the initial volume spanned
by the nearest-neighbor atoms is much smaller for $V_{\rm P}$ than for $V_{\rm
Ga}$. As a consequence, the relaxation results in an {\em increase} of the
surface buckling near the defect ($0.85$\,\AA{} as compared to $0.56$\,\AA{})
and not in a decrease as for $V_{\rm Ga}$.  

The distances between the atoms surrounding the defect are $2.8$\,\AA{} for
the $\rm Ga_1Ga_2$ and $3.7$\,\AA{} for $\rm Ga_1Ga_1$ separ\-ation. 
This compares to $3.8$\,\AA{} in bulk GaP and $2.5$\,\AA{} in bulk
gallium.~\cite{land}  
The angle between the back bonds of the surface gallium atoms $\rm Ga_1$ is 
$120^{\rm o}$ as compared to $122^{\rm o}$ for the relaxed unperturbed
surface.
Thus, like the cations at the surface without a vacancy, the atoms $\rm Ga_1$
prefer an $sp^2$ bonding configuration.
While this is the first {\it ab initio} calculation of anion surface vacancies
at GaP (110), there have been several publications for anion vacancies at the
(110) surface of GaAs: 
Calculating $V_{\rm As}$ in a $(2\times 2)$ surface cell, Yi {\it et
al.}~\cite{yel_2:95} determined the height difference between surface gallium
atoms for the neutral vacancy to be  $0.2$\,\AA{} and for the negatively
charged defect to be $0.4$\,\AA{}. 
Zhang and Zunger~\cite{zhang:96-1} calculated the inward relaxation of the atom
$\rm Ga_1$ towards $V_{\rm As}$ to be $0.34$\,\AA{} and $0.29$\,\AA{} in the
negative and positive charge state.
The positively charged vacancy at GaAs (110) has also been calculated by Kim and
Chelikowsky,~\cite{cheli:96} who give an inward relaxation of $0.25$\,\AA{}.
Keeping in mind that GaP has a slightly smaller lattice constant than GaAs,
these values are in close agreement with our result of $0.3$\,\AA{}.

The analysis of the electronic structure reveals also similarities between
cation and anion surface vacancies. 
Again, we find two defect levels in the fundamental band gap
[Fig.\,\ref{fig02}(d)].  
The level $1a^{\prime\prime}$ is located 0.8\,eV above the valence-band
maximum and the level $2a^{\prime}$ mixes with the unoccupied surface
bands close to the conduction-band minimum. 
The occupation numbers are again one and zero for these defect states.
In the valence band we find two localized states at about $-0.8$\,eV and
$-6.5$\,eV. 
Since these are resonances in the valence band, they are not shown in
Fig.\,\ref{fig02}(d).
The levels  $1a^{\prime\prime}$  and $2a^{\prime}$  consist mainly of
$p$-states centered at the gallium neighbors, while the lower-lying resonance
is mostly $s$-like in character.
As for $V_{\rm Ga}$, the energeti\-cally lowest level is
symmetric with respect to the surface mirror plane, $1a^{\prime\prime}$ is
antisymmetric, and $2a^{\prime}$ is symmetric. 
The electronic structure of the anion surface vacancy has been previously
calculated by Daw and Smith~\cite{daw:79,daw:81} using a tight-binding method
combined with Green's-function techniques. 
Without considering atomic relaxations, they found the same sequence of
electronic states for GaAs,~\cite{daw:79} namely a symmetric level near the
valence-band maximum and two levels in the band gap, where the lower,
antisymmetric level is below midgap and the symmetric level is  near the bottom
of the conduction band.
In the calculations for GaP,~\cite{daw:81} the authors determined the position
of the highest occupied level to be at $1.0$\,eV, which is in good agreement
with our value of $1.1$\,eV for the {\em unrelaxed} surface vacancy.  
In contrast to these results, tight-binding calculations by Allen, Sankey, and
Dow~\cite{all:86} predicted  three deep levels at $1.1$, $1.5$, and
$2.2$\,eV. 

Comparing our results with recent {\it ab initio} calculations for GaAs (110), we
find qualitative agreement for the position and character of levels: 
Kim and Chelikowsky~\cite{cheli:96} obtained for the positively charged
vacancy the same symmetry of the vacancy states and
very similar positions of levels to those we find for the neutral vacancy at GaP
(110). 
Yi {\it et al.}~\cite{yel_2:95} determined the half-filled level above
the center of the band gap while Zhang and Zunger~\cite{zhang:96-1} calculated
this level to be $0.41$ and $0.5$\,eV above the valence-band maximum for the
positive and negative charge state. 

Similar to the gallium surface vacancy, we expect also for the
phosphorus surface vacancy at least two charge-transfer levels corresponding
to the possible charge states ($+,0,-$).
This result is consistent with STM measurements by Ebert {\it et
al.},~\cite{ebert:94-1,ebert:93-4} who determined the phosphorus surface
vacancy to be positively charged on $p$-type material and neutral or
negatively charged under $n$-type conditions.

\section{Subsurface vacancies}\label{section4}

\begin{table}
\caption[]
 {\label{tab03}
Relaxations of the nearest-neighbor atoms of the cation ($V_{\rm Ga}$) and
anion ($V_{\rm P}$) vacancy in the third surface layer and in the 64-atom
bulk supercell. 
For the vacancies in the third surface layer, the relaxations for the nearest
neighbors in the second, third, and fourth layer are given in percent of the
bulk bond distance ($2.32$\,\AA{}). For the bulk vacancies, the
breathing mode relaxation  maintaining the $T_{ d}$ symmetry is
indicated.
All values are given in percent of the bulk bond distance ($2.32$\,\AA{}). \\
}
\begin{tabular}{c*{3}{r@{.}l}*{1}{r@{.}l}}
 Vacancy     &\multicolumn{6}{c}{Subsurface} & \multicolumn{2}{c}{Bulk}\\
             & \multicolumn{2}{c}{layer 1} & \multicolumn{2}{c}{layer 2} &
             \multicolumn{2}{c}{layer 3} &\multicolumn{2}{c}{\ }    \\
\hline
$V_{\rm Ga}$ & 7&{0}   & 2&{1} & 4&{5} & 3&{8}  \\
$V_{\rm P}$  & $-14$&{0} & 8&{8} & 5&{1} & 6&{1} \\
\end{tabular}
\end{table}

Defects at and near the surface differ from their corresponding bulk defects
due to changes in coordination number as well as in symmetry.
In order to compare defects at different positions relative to the surface in
a systematic way, we have calculated both types of  vacancies in a bulk system
as well as in the second and third layer of a surface cell.
We will first focus on bulk defects and vacancies in the third
layer. Vacancies in the second layer will be discussed separately: They
exhibit features observed neither for bulk defects nor for surface defects.

\subsection{Bulk defects}\label{section4a}
Bulk vacancies in a zinc-blende semiconductor are characterized by a bonding
level $a_1$ and a threefold-degenerate level
$t_2$.~\cite{bach:81,scheff:81,bar:85,scheff:84,scheff:88,scheff:93,laa:92,seit:94,north:94}
For vacancies in the neutral charge state, the $a_1$ level lies in the valence
band and the $t_2$ states are located above the valence-band maximum for the
cation vacancy and near the conduction-band minimum for the anionic defect.
Table\,\ref{tab03} summarizes our results for structural changes and
Figs.\,\ref{fig02}(c) and \ref{fig02}(f) show the position of levels 
for the relaxed defects. 
Atomic relaxation is mainly governed by an inward breathing relaxation
conserving the tetrahedral symmetry of the lattice. 
We find the inward relaxation of the gallium vacancy to be 3.7\% of the bulk
bond distance ($2.32$\,\AA{}).  
For the unrelaxed vacancy the $a_1$ resonance is located $-1.0$\,eV below the
valence-band maximum as compared to $-0.75$\,eV calculated by Scheffler {\it
et al.}~\cite{scheff:81} with a self-consistent pseudopotential Green's-function 
method. The $t_2$ level lies at $0.2$\,eV, in fair agreement with
Ref.~\onlinecite{scheff:81} ($0.15$\,eV).  
Allowing for atomic relaxation, the $a_1$ level shifts to $-1.7$\,eV while the
$t_2$ level gains $0.1$\,eV.

Calculating the atomic relaxations of $V_{\rm P}$, we find an inward relaxation
of 6.1\% (Table\,\ref{tab03}).
For the ideal atomic positions, the $a_1$ resonance is found at about $0.1$\,eV
below the valence-band maximum.
Including atomic relaxations lowers the energy by $0.4$\,eV.
The $t_2$ level is found within the conduction band for the unrelaxed defect
and lies below the band edge after relaxation.

An analysis of our bulk and slab calculations shows that these results are
almost insensitive to whether the vacancies are in a bulk cell or in the third
layer of a surface cell. Both the positions of the defect levels $a_1$ and
$t_2$ and the formation energies (which will be discussed in
Sec.\,\ref{section6}) agree well. The main differences are slight
modifications in the atomic geometry. The origin of this is the changes in
symmetry due  to the presence of the surface (Table\,\ref{tab03}).

\subsection{Defects in the second layer}\label{section4b}

\begin{figure}[t]
\centerline{\psfig{file=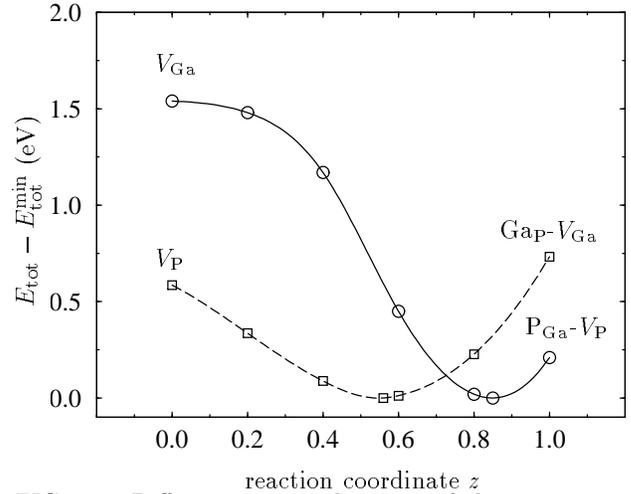,height=6.5cm,clip=t}}
\caption[]
 {\label{fig03}
Differences in total energy of the vacancies in the second layer and the
complexes of surface vacancy and antisite in the second layer. 
The reaction coordinate $z$ is the lateral position of the nearest-neighbor
atom $\rm P_1$ ($\rm Ga_1$) of $V_{\rm Ga}$ ($V_{\rm P}$) as defined in
Fig.~\ref{fig01}\,(c)\,(d).  
At $z=0.0$, this atom is confined to the plane of the surface anions (cations),
and at  $z=1.0$ it is placed in the plane of the cations (anions) in the
second layer. 
Solid line, $V_{\rm Ga}$ and $\rm P_{\rm Ga}$-$V_{\rm P}$;
dashed line, $V_{\rm P}$ and $\rm Ga_{\rm P}$-$V_{\rm Ga}$.
}
\end{figure}

\begin{table}
\caption[]
 {\label{tab04}
Relaxations of the nearest-neighbor and two second nearest-neighbor atoms of
the cation ($V_{\rm Ga}$) and anion ($V_{\rm P}$) vacancies in the second
surface layer. $A_x$ refers to the atom of species $A$ in the surface
($x\,=\,1$) or second and third surface layer ($x=\,2,3$). 
 $\Delta{\rm [001]}$, $\Delta{\rm [1\overline{\rm 1}0]}$, and $\Delta{\rm [110]}$
 indicate the shifts in the directions illustrated in
 Fig.\,\ref{fig01}, and $||\Delta||$ is the total displacement.
All values are given in \AA{}.\\
}
\begin{tabular}{cc*{5}{r@{.}l}}
     &   &\multicolumn{8}{c}{$V_{\rm Ga}$} \\ 
     &   &\multicolumn{2}{c}{$\rm P_1$} & \multicolumn{2}{c}{$\rm P_2$} &
     \multicolumn{2}{c}{ $\rm P_3$} &\multicolumn{2}{c}{ $\rm Ga_1$}
     &\multicolumn{2}{c}{ $\rm Ga_2$}  \\
\hline
$\Delta{\rm [001]}$       &  
& 0&{47}&$-0$&{06}& $-0$&{10} &0&{21}&0&{09}  \\
$\Delta{\rm [1\overline{\rm 1}0]}$ &  
&\multicolumn{2}{c}{  }& 0&{09} &\multicolumn{2}{c}{  }&$-0$&{03}&0&{02} \\
$\Delta{\rm [110]}$       &  
&$-1$&{70}&$-0$&{03}&0&{08} &$-0$&{40}&$-0$&{03} \\
 \\
 $||\Delta||$       & 
& 1&{76}& 0&{11}&0&{13} & 0&{45}& 0&{10} \\
\\ \\
     &   &\multicolumn{8}{c}{$V_{\rm P}$} \\ 
     &   &\multicolumn{2}{c}{$\rm Ga_1$} & \multicolumn{2}{c}{$\rm Ga_2$} &
     \multicolumn{2}{c}{ $\rm Ga_3$} &\multicolumn{2}{c}{ $\rm P_1$}
     &\multicolumn{2}{c}{ $\rm P_2$}  \\
\hline
$\Delta{\rm [001]}$       &  
&$-0$&{58}& 0&{01}&  0&{02} &$-0$&{41}&$-0$&{05}  \\
$\Delta{\rm [1\overline{\rm 1}0]}$ &  
&\multicolumn{2}{c}{  }& 0&{06}&\multicolumn{2}{c}{  }& 0&{21}& 0&{02} \\
$\Delta{\rm [110]}$       &  
&$-0$&{85}& 0&{03}&  0&{02} &$-0$&{28}&$-0$&{05}  \\
 \\
 $||\Delta||$       & 
& 1&{03}& 0&{07}&  0&{03} & 0&{54}& 0&{07}
\end{tabular}
\end{table}

Let us now focus on neutral vacancies in the second layer. The results for the
equilibrium structure are summarized in Table\,\ref{tab04} and
Figs.\,\ref{fig01}(c) and \ref{fig01}(d).
As for surface vacancies, the atomic structure of cation and anion
vacancies maintains the $C_{1h}$ symmetry of the defect-free surface.
Analyzing the atomic relaxations, we find significant displacements for the
nearest-neighbor atom at the surface.
For both defects this atom moves almost into the center of the
vacancy. Therefore, vacancies in the second layer could also be considered as
vacancy-antisite complexes  (${\rm P_{\rm Ga}}$\,-\,$V_{\rm P}$ and ${\rm
Ga_{\rm P}}$\,-\,$V_{\rm Ga}$).  
In contrast to previous calculations for {\em bulk} vacancies 
(e.g., $V_{\rm Ga}$ and ${\rm As_{\rm Ga}}$\,-\,$V_{\rm As}$ in 
GaAs),~\cite{bar:85a,bar:86,bock:97} we do not find the {\em second-layer
  subsurface} vacancies to be metastable.
Figure~\ref{fig03} shows the energy of the defects as a function of the
$z$-coordinate of the nearest-neighbor atom  $\rm P_1$ ($\rm Ga_1$) at the
surface.  
At $z=0.0$ this atom is confined to the plane of the surface anions (cations),
and at  $z=1.0$ it is placed in the plane of the cations (anions) in the
second layer. 
In the equilibrium structure, we find the total displacement of the atom $\rm
P_1$ next to $V_{\rm Ga}$ to be  $1.8$\,\AA{}, and that of the atom $\rm
Ga_1$ next to $V_{\rm P}$ is  $1.0$\,\AA{}. 
These values can be compared to $2.3$\,\AA{} for the bond length between
atoms in the first and second layer on the relaxed unperturbed surface.
In contrast to the surface vacancies, we find a substantial displacement also
for the second nearest-neighbor atoms in the surface layer.
The two nearest-neighbor atoms in the second and the one in the third layer of
the slab undergo almost no relaxation (Table\,\ref{tab04}).

The gallium vacancy has two levels within the fundamental band gap
[Fig.\,\ref{fig02}(b)]: one doubly occupied at $0.6$\,eV and one singly occupied
at about $1.2$\,eV.  
These states are antisymmetric and symmetric with respect to the surface mirror.
For the phosphorus vacancy, we find a symmetric and  singly occupied
defect state at the middle of the band gap and an empty and antisymmetric
state below the conduction-band minimum [Fig.\,\ref{fig02}(e)]. 
As for surface vacancies, we expect the cation {\em and} the anion vacancy
in the second layer to show an amphoteric behavior: While under $p$-type
conditions they should act as a donor, $n$-type conditions might stabilize a
negatively charged state. 
Thus cation and anion vacancies in the first and second layer are possible
candidates for pinning the surface Fermi level.

We conclude that the equilibrium structure of subsurface vacancies
is mainly determined by the coordination of the nearest-neighbor atoms: 
Twofold-coordinated atoms in the surface layer can alter their position
signifi\-cantly without breaking the remaining bonds. 
However, threefold-coordinated atoms as completely surrounding a vacancy in
the third or a deeper layer are strongly bound to their positions in the
unperturbed lattice.  
Thus the ability of atoms in the second layer to relax is comparable to that
of bulk atoms. Consequently, vacancies with only threefold neighbors 
(i.e., vacancies below the second layer) show very similar properties to bulk 
defects, while vacancies that have twofold-coordinated neighbors
(i.e., vacancies at the surface and in the second layer) show a very different
geometry and electronic structure.

\section{Defect formation energies}\label{section6}

\begin{table}
\caption[]
 {\label{tab05}
Formation energies for surface and subsurface vacancies under phosphorus-rich
conditions. The table shows the results for unrelaxed and relaxed defects per
broken bond in order to compare the threefold-coordinated surface vacancies to
the fourfold-coordinated subsurface and bulk defects. All values are given in
eV.\\
}
\begin{tabular}{lcccc}
          &\multicolumn{4}{c}{$V_{\rm Ga}$} \\
          & surface & second layer & third layer & bulk \\
\hline
unrelaxed &0.99 &1.06 &1.06 &1.02  \\
relaxed   &0.65 &0.55 &1.01 &0.97  \\
\\
\\
 & \multicolumn{4}{c}{$V_{\rm P}$} \\
          & surface & second layer & third layer & bulk \\
\hline
unrelaxed &1.05 &1.05 &1.08 &1.09  \\
relaxed   &0.82 &0.81 &1.03 &1.05  
\end{tabular}
\end{table}

In the preceeding sections we have focused on the atomic geometry and  the
electronic structure of the defects, and how these properties are modified if
a bulk defect is brought to the surface. We will now discuss how the energy
required to create a vacancy in thermodynamic equilibrium (i.e., the formation
energy) is affected by the surface. 
The formation energy of a defect is given by 

\begin{equation}\label{form1.eqn}
  E_{\rm f} =  E_{\rm tot}^{\rm defect} - N_{\rm Ga} \mu_{\rm Ga}
  - N_{\rm P} \mu_{\rm P} - q  E_{\rm Fermi}
 \mbox{\ \ \ .}
\end{equation}

\noindent
$N_{\rm Ga}$ and $N_{\rm P}$ are the number of cations and anions with
chemical potentials $\mu_{\rm Ga}$ and $\mu_{\rm P}$, and  $q$ is the number
of excess electrons in exchange with the electron reservoir $E_{\rm Fermi}$. 
Since we are focusing on neutral defects ($q\!=\!0$), the formation energies are
independent of the Fermi level. 
The Ga and P chemical potentials are related to each other by the condition of
chemical equilibrium with the bulk GaP crystal: 

\begin{equation}
 \mu_{\rm GaP}^{\rm bulk} = \mu_{\rm Ga} + \mu_{\rm P}
 \mbox{\ \ \ .}
\end{equation}

\noindent 
The formation energy thus can be written as~\cite{scheff:93}

\begin{equation}\label{e_bild}
 E_{\rm f}
   = E_{\rm tot}^{\rm defect} -\mu_{\rm P}(N_{\rm P}-N_{\rm Ga}) 
   - N_{\rm Ga}\mu_{\rm GaP}^{\rm bulk} 
 \mbox{\ \ \ .}
\end{equation}

\noindent
Moreover, we can approximate the upper limit of the chemical potential of each
species by their corresponding bulk phases ($ \mu_{\rm Ga}^{\rm bulk}\mbox{, }
\mu_{\rm P}^{\rm bulk}$). 
With the definition of the heat of formation of the crystal,

\begin{equation}
 \Delta H_{\rm f} = \mu_{\rm GaP}^{\rm bulk} - \mu_{\rm Ga}^{\rm bulk} - 
   \mu_{\rm P}^{\rm bulk}
 \mbox{\ \ \ ,}
\end{equation}

\noindent
we get a range in which the chemical potential can be
varied from gallium-rich to phosphorus-rich conditions:

\begin{equation}\label{mu_borders3.eqn}
 \mu_{\rm P}^{\rm bulk} - \Delta H_{\rm f} \le  \mu_{\rm P} 
     \le  \mu_{\rm P}^{\rm bulk} 
 \mbox{\ \ \ .}
\end{equation}

\noindent
Assuming metallic orthorhombic gallium~\cite{land,tos:95-1} and black
phosphorus in an orthorhombic unit cell~\cite{gmel:64} to be the
equilibrium structures of the condensed elemental phases, we get
$\Delta H_{\rm f}=1.13$\,eV. This value is slightly higher than the experimental
result ($1.08$\,eV).~\cite{bech:88} 

\begin{figure}[t]
\centerline{\psfig{file=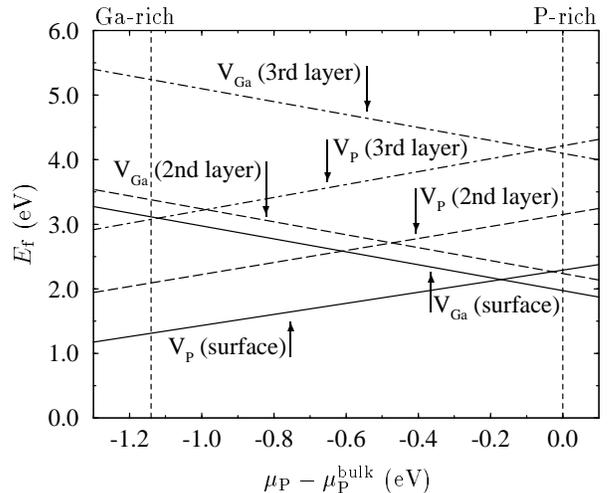,height=6.5cm,clip=t}}
\caption[]
 {\label{fig04}
Formation energies for relaxed surface and subsurface vacancies as a function
of the difference of the chemical potential of phosphorus and bulk phosphorus.
The limits for Ga-rich ($\mu_{\rm P}-\mu_{\rm P}^{\rm bulk}=-1.13$\,eV) and
P-rich conditions ($\mu_{\rm P}-\mu_{\rm P}^{\rm bulk}=0.00$\,eV) are defined in
the text. 
Note that the defects labeled as $V_{\rm Ga}$ (second layer) and $V_{\rm P} 
$(second layer) are better described as ${\rm Ga_{\rm P}}$\,-\,$V_{\rm P}$ and  
${\rm P_{\rm Ga}}$\,-\,$V_{\rm Ga}$ complexes.
}
\end{figure}

Figure\,\ref{fig04} summarizes our results for the different relaxed defects.
Over a wide range of the chemical potential, the anion surface vacancy is the
energetically most favored defect. Only under very phosphorus-rich conditions
is the cation surface vacancy preferred. 
Vacancies in the sec\-ond and third layer show a significantly higher formation
energy. Creating a gallium vacancy in the sec\-ond layer takes $0.25$\,eV more
than at the surface while the other subsurface vacancies are even less
favored.

Table\,\ref{tab05} lists the formation energies with and without considering
atomic relaxations. Note that all energies in Table\,\ref{tab05} are given per
broken bond and not per defect. This allows us to compare directly bulk and
surface vacancies that have different coordinations. 
For simplicity, all formation energies are given under phosphorus-rich
conditions. 
According to Eq.~(\ref{e_bild}), the following discussion remains
unaffected by considering a different chemical environment. 
Let us first discuss the case without any atomic relaxation: The most striking
feature is the fact that the energy per broken bond is almost unchanged with
respect to the position of the vacancy (surface, subsurface,
bulk).~\cite{foot:2}  
The deviations are smaller than $0.1$\,eV, i.e., close to the estimated error of
our method.
The largest difference ($70$\,meV) is found for $V_{\rm Ga}$ between the first
and sec\-ond surface layer. 

However, if we consider atomic relaxation, the picture changes qualitatively:
For vacancies in the first {\it and} sec\-ond layer, we find a strong relaxation. 
Due to this relaxation the formation energy is largely reduced (up to
$0.5$\,eV, i.e., the energy gained by atomic relaxation is one order of 
magnitude larger than for bulk vacancies). 
It is further interesting to note that the relaxation energy for $V_{\rm Ga}$
is about twice as large as for $V_{\rm P}$. 

For bulk vacancies and vacancies in the third surface layer, the lowering of 
the formation energy due to atomic relaxation is small and almost independent of
the type of defect.  
It has been shown in Sec.\,\ref{section4a} that these defects are very
similar with respect to defect levels and atomic equilibrium positions. The
presence of the surface also induces differences in formation energy that are
less than $40$\,meV indicating the bulklike nature of the subsurface vacancies.

\section{Simulation of STM images}\label{section5}
\subsection{Method}
For simulating STM images we employ the Tersoff-Hamann
approach.~\cite{bar:61,ter:85} 
In this approach the tunneling current is proportional to the local density
of states at the tip position, $\rho_{s}({\bf r^{\rm tip}},\epsilon)$,
integrated over the interval between the Fermi surface and the applied bias $U$: 

\begin{equation}\label{tersoff-hamann0}
  I({\bf r^{\rm tip}},U)~\propto~ \int_{E_{\rm Fermi}}^{E_{\rm Fermi}+eU}
           \rho_{s}({\bf r^{\rm tip}},\epsilon)\  d\epsilon              
      \mbox{\ \ \ .}
\end{equation} 

A positive bias results in imaging the empty states in the specified energy
range at position ${\bf r^{\rm tip}}$ and nega\-tive bias in imaging the filled
states. 
By mapping surfaces of constant $I({\bf r^{\rm tip}},U)$ an isocurrent
picture can be simulated.  
Alternatively the integration might be evaluated for fixed tip height.  
Calculations of various reconstructed semiconductor
surfaces~\cite{feen:87-2,wil:92,dab:95,ebert:95-2} as well as surfaces with
impurities and adsorbates~\cite{zhang:96-1,cheli:96,lim:95,jan:95} have proven
this approach to be very useful for the interpretation of experimental
results.

\begin{figure}[t]
\centerline{\psfig{file=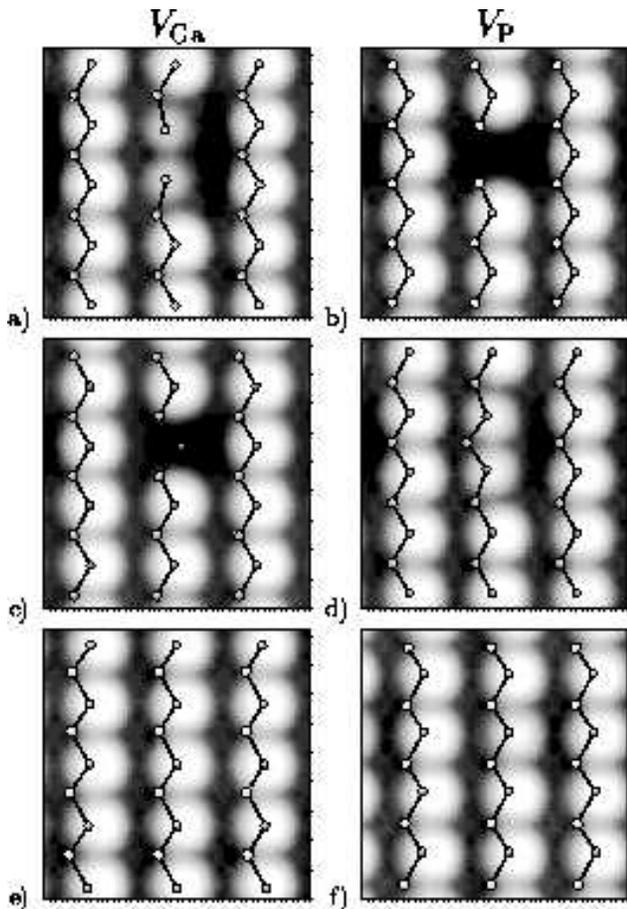,height=12cm,clip=t}}
\caption[]
 {\label{fig05}
STM simulations of occu\-pied states for surface and subsurface vacancies. 
(a)~Gallium surface vacancy; 
(b)~Phosphorus surface vacancy; 
(c)~$V_{\rm Ga}$ in the second layer; 
(d)~$V_{\rm P}$ in the second layer; 
(e)~$V_{\rm Ga}$ in the third layer; 
(f)~$V_{\rm P}$ in the third layer.
The electron density in the plots varies from
$0.4\times10^{-7}\mbox{\AA{}}^{-3}$ to 
$4.0\times10^{-7}\mbox{\AA{}}^{-3}$ and the tip surface separation is
estimated to be $3$\,\AA{}. 
}
\end{figure}

\begin{figure}[t]
\centerline{\psfig{file=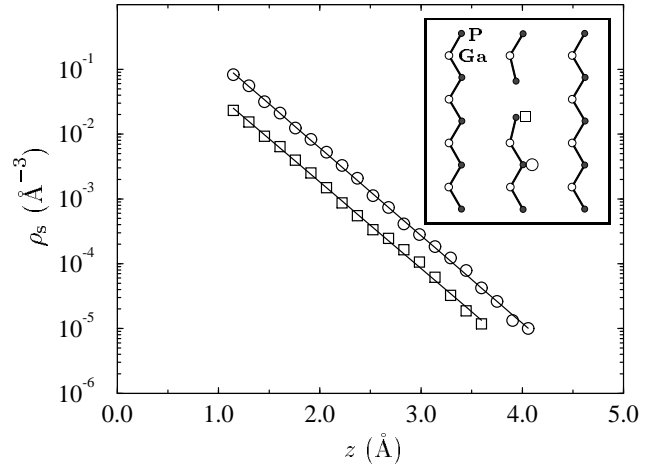,height=6.5cm,clip=t}}
\caption[]
 {\label{fig06}
The local density $\rho_{s}$ as a function of the tip height above the
surface.
The inset shows the same part of the surface as Fig.\,\ref{fig05}(a) (gallium
surface vacancy). Squares correspond to the position of the nearest-neighbor
surface dangling bond and circles to that of the next anion in the direction of
the zigzag row. 
The decay constant deduced from fitting the data is $3.1\,\rm \AA^{-1}$ for
both graphs and the height difference for constant density is $0.42$\,\AA{}.
As in Fig.\,\ref{fig05}, $\rho_{s}$ includes all occupied defect and
surface states as well as surface resonances down to $1.4$\,eV below the
valence-band maximum. 
The zero of the abscissa corresponds to the position of the outermost anions
at the relaxed 
ideal surface. 
}
\end{figure}

\subsection{Results}
Figures\,\ref{fig05}(a) and \ref{fig05}(b) display our results for the occupied states of the
gallium and phosphorus surface vacancy. The plots were taken at a distance of
3~\AA{} from the outermost anion layer and include all occupied defect and
surface states as well as surface resonances down to $1.4$\,eV below the
valence-band maximum.  
The pictures are dominated by the filled dangling bonds above the anions. 
Thus, for the defect-free surface the STM picture shows the period\-icity of
the surface lattice and the maxima of the charge density are close to the 
positions of the surface anions. 
  
If we focus on the cation vacancy [Fig.\,\ref{fig05}(a)]
we find a localized perturbation in the lattice: the dangling bonds of the 
two neighboring atoms along the $[1\overline{\rm 1}0]$ direction are laterally
displaced according to the relaxation of the atoms and clearly appear less
intense than the ones farther away from the defect.  
This signature of the defect has also been identified in STM
experiments.~\cite{ebert:93-4,ebert:98-1}  
However, from the experimental data it is not clear whether 
this effect is due to  atomic relaxation or due to changes in the electronic
structure near the vacancy.  
We therefore compared the LDOS above the surface dangling bonds of the
neighboring atoms with that of dangling bonds farther away from the defect. 
The results are shown in Fig.\,\ref{fig06} together with an inset defining the
lateral positions where the tip height has been varied. 
In the investigated interval above the surface we find an  exponential decay
of the density into the vacuum.  
The decay constants at the two maxima are virtually identical
($3.1\,\rm\AA^{-1}$).   
We therefore expect that  in constant current STM the height difference
between the two maxima is not a function of the tunneling current. 
From Fig.\,\ref{fig06} we find this height difference to be $0.42$\,\AA{},
which is very close to the difference in the atomic structure  ($0.44$~\AA{},
Table\,\ref{tab02}).  
We note that the good agreement between the height corrugation in the
electronic and atomic structure is related to the specific structure of the
(110) surface and that of the defect: Both rehybridization and charge transfer
for this system are small, and our STM simulation mainly shows the filled
surface dangling bonds, which are not significantly changed by the presence of the
vacancy. This is true even for nearest-neighbor atoms.  
We expect larger deviations for charged surface defects or for more complex
systems where defect states are less localized or where rehybridization of
atoms near the defect is more prominent.

The STM simulation of the anion surface vacancy is characterized by a
depletion of the density of states due to the missing dangling bond
[Fig.\,\ref{fig05}(b)]. 
The perturbation is very localized above the defect site and already the
dangling bonds located at the second nearest neighbors are not influenced by
the defect. 
This agrees well with experimental data for the neutral
defect~\cite{ebert:94-1} and with the short-range character of the defect's
perturbation of the surface as discussed in Sec.\,\ref{section3}. 

Simulating the  vacancies in the second layer, we get the results shown in
Figs.\,\ref{fig05}(c) and \ref{fig05}(d). Due to the large relaxation of the 
anion in the surface
layer, the gallium vacancy displays features similar to the surface phosphorus
vacancy: The phosphorus atom above the vacancy has no states within the
simulated energy range at that location, which can be explained by the large
vertical displacement of this atom.  
For the phosphorus vacancy the dangling bonds of the second nearest neighbors
are slightly displaced and depressed. 
As for the cation surface vacancy, we analyzed this behavior as a function
of the tip height. We find the height difference of the tip to be $0.22$~\AA{} 
as compared to $0.27$~\AA{} structural relaxation.  
Thus, also the signature of this vacancy is mainly governed by the relaxation
of the surface atoms. 
We note that the STM simulations of the gallium surface vacancy
[Fig.\,\ref{fig05}(a)] and of the phosphorus vacancy in the second layer
[Fig.\,\ref{fig05}(d)] show a remarkable resemblance. 
The same holds for the anion surface vacancy [Fig.\,\ref{fig05}(b)] and the
cation vacancy in the second layer [Fig.\,\ref{fig05}(c)].
Therefore,  we expect these defects  to be hardly  discernible in STM
measurements of the occupied states. 
However, for the  gallium vacancy at the surface and $V_{\rm P}$ in the second 
layer, the highest occupied defect state has a different symmetry
[Figs.\,\ref{fig02}(a) and \ref{fig02}(e)]. Thus, both defects should be 
distinguishable in
experiment. Also $V_{\rm P}$ at the surface and  $V_{\rm Ga}$ in the 
second layer should be distinguishable in experiments such as  scanning tunneling
spectroscopy because they exhibit very different positions of the highest
occupied defect state [Figs.\,\ref{fig02}(b) and \ref{fig02}(d)]. 

The simulation of a gallium vacancy in the third layer is virtually identical
to that of a defect-free surface [Fig.\,\ref{fig05}(e)]. For the phosphorus
vacancy, a slight depression above the missing atom is found
[Fig.\,\ref{fig05}(f)], which, however, might be too weak to be resolved in
experiment. We therefore expect that {\em neutral} vacancies below the second
surface layer cannot be detected  by STM measurements. 
For {\em charged} defects the local surface band bending induced by the
impurity might change  this result
qualitatively.~\cite{john:93e,ebert:96-4}

\section{Summary}
Using first-principles total-energy calculations we investigated neutral
surface and subsurface vacancies on both sublattices at GaP (110).  
Atomic relaxations, electronic structure, as well as formation energies were
analyzed, and  STM pictures of the various vacancies were simulated.

For both the gallium and the phosphorus surface vacancy, we find a pronounced
inward  relaxation of nearest neighbors into the defect. The electronic
structure is characterized by a singly occupied deep level for $V_{\rm Ga}$
and $V_{\rm P}$.
Thus we expect them to be amphoteric, indicating that surface vacancies are
electrically active and potentially a source of compensation.
From the third surface layer on, the defects show essentially the same
properties as bulk vacancies, namely a breathing-mode-like relaxation of the
nearest-neighbor atoms and a triply degenerate acceptor ($V_{\rm Ga}$)
and donor level ($V_{\rm P}$), respectively. 
Vacancies in the second surface layer are found to be unstable against the
formation of vacancy antisite complexes where the nearest-neighbor surface
atom moves towards the vacancy site. 
For $V_{\rm Ga}$, the highest occupied level lies near the empty surface band
while for  $V_{\rm P}$ it is found at a midgap position.

As can be expected from their lower coordination number, surface vacancies are 
energetically preferred as compared to subsurface and bulk vacancies. 
Our results show that, especially at and near the surface, the formation energy
of defects is significantly affected by rather large atomic relaxations.
For all vacancies with nearest-neighbor atoms in the surface layer
we find relaxation energies that are at least six times higher than for
defects with four threefold-coordinated nearest-neighbor atoms.

Our calculated STM images of the surface vacancies agree well with available
experimental data.  
The comparison of the STM simulation with the atomic structure shows that 
{\em neutral} vacancies at this surface exhibit a close relation between
atomic relaxation and height corrugation in the STM picture.
To our knowledge, unlike charged subsurface defects, which have been frequently
observed using STM, uncharged subsurface vacancies have not been reported by
experimental groups so far.
This is consistent with our results for vacancies from the third layer on,
which exhibit virtually no interaction with the surface. 
However, vacancies in the second layer look like surface vacancies in our
simulation. In experiment they might be discerned from surface vacancies by
spectroscopic methods. 

\begin{acknowledgments}
The authors would like to thank Ph. Ebert for stimulating discussions and
unpublished data. 
This work was supported by the Deutsche Forschungsgemeinschaft.
\end{acknowledgments}



\begin{thebibliography}{10}
\bibitem{foot:4}
Following the notation for the irreducible representation of the $C_{ 1h}$
point group, we label the states $a_{n}^{\prime}$ ($a_{n}^{\prime\prime}$) if
they are symmetric (antisymmetric) with respect to the surface mirror plane.
\bibitem{foot:2}
Note that the relative formation energy of cationic and anionic vacancies
  depends on chemical potential as illustrated in Fig.~(\ref{fig04}).
\end{thebibliography}
\end{document}